# Seismic analysis based on a new interval method with incomplete information

Shizhong Liang, Yuxiang Yang, Chen Li, Feng Wu[*]

*State Key Laboratory of Structural Analysis of Industrial Equipment, Department of Engineering Mechanics, Dalian University of Technology, Dalian, Liaoning Province, 116024, P.R. China*

[*]Corresponding author: Tel: 8613940846142; E-mail address: wufeng_chn@163.com

May 2025

**Abstract**

For seismic analysis in engineering structures, it is essential to consider the dynamic responses under seismic excitation, necessitating the description of seismic accelerations. Limit seismics samples lead to incomplete uncertainty information, which is described by the non-probabilistic method reasonable. This study employs the minimum interval radius-based interval process (MRIP) based on the convex model to describe the time-variant uncertain seismic acceleration, subsequently conducting uncertainty analysis for seismic structures. However, the Monte Carlo simulation for uncertainty analysis requires extensive deterministic computations to ensure accuracy, exhibiting poor computational efficiency. To address this issue, this paper first improves the covariance matrix adaptation evolution strategy (CMA-ES) through the dynamic evolution sequence, proposing DES-ES, whose efficiency is validated to be higher than that of CMA-ES. Furthermore, leveraging the dependency of the responses, a computational framework named DES-ES-SS is proposed. Numerical experiments demonstrate that DES-ES-SS improves computational efficiency while maintaining the accuracy of the interval uncertainty analysis of the seismic structures whether the seismic acceleration is stationary or non-stationary.

**Statements and Declarations**

Competing Interests: The authors declare that they have no known competing financial interests or personal relationships that could have appeared to influence the work reported in this paper.

**Acknowledgments**

The authors are grateful for the support of the Natural Science Foundation of China (No. 12002396), and the Basic Science Center Program of the National Natural Science Foundation of China (No. 62388101).

# 1 Introduction

For seismic analysis of engineering structures, the description of the time-variant uncertain seismic acceleration is critical. Currently, seismic acceleration is often described by the stochastic process [1]. Like wind [2] and waves [3], seismic acceleration acts as an environmental load, with its frequency-domain properties defined by power spectral density (PSD) or evolutionary PSD (EPSD) [4, 5]. By analyzing Kanai's research[6], Tajimi [7] proposed the Kanai-Tajimi spectrum describing seismic acceleration, which exaggerates the energy in the low-frequency part. To solve the problem, Clough and Penzien [8] proposed the Clough-Penzien spectrum by introducing a high-pass filter. At present, Zhang et al. [9, 10] have further used the frequency domain information to design new seismic metamaterials. Once the PSD or EPSD is constructed, samples can be obtained via techniques like the Karhunen-Loève (K-L) expansion [11] and the stochastic harmonic function method [12], followed by uncertainty analysis in the time domain [13] using methods such as the Monte Carlo simulation (MCS) [14], the quasi-MCS [15, 16], and the perturbation method [17, 18].

However, accurate uncertainty analysis requires the construction of precise stochastic processes. When the uncertainty information is incomplete, the constructed model is inaccurate [19]. For seismic acceleration, the samples are difficult to obtain, resulting in incomplete information. Therefore, it is appropriate to use non-probabilistic methods for description [20]. Jiang et al. [21] proposed the interval process (IP), which uses the convex model to describe time-variant uncertainty. Furthermore, Ni et al. [22] proposed an efficient sampling method for IP, the interval K-L expansion. On this basis, Wu et al. [23] proposed the minimum interval radius-based IP (MRIP) and the efficient construction method. In this paper, MRIP is used to describe the seismic acceleration with incomplete information, and then conduct uncertainty analysis on the seismic structure. Since MRIP is a non-probabilistic method, it only involves interval information. Therefore, when using MRIP to describe seismic acceleration, it is only necessary to solve the envelope interval of the response, i.e., to conduct interval uncertainty analysis.

For the interval uncertainty analysis of systems with time-variant uncertainties described by IP, a large number of samples obtained by the interval K-L expansion can be used to obtain the envelope interval of the response through a large number of deterministic calculations, i.e., MCS for interval uncertainty analysis. The inherent accuracy-efficiency tradeoff in MCS requires extensive sampling for precise results, leading to extremely low efficiency. To address the issue, Ni et al. [24] proposed the sequential simulation strategy by utilizing the dependency of the response. Wu et al. [25] regarded the solution of the interval of the time-invariant response as extreme-value problems and then used one of the most advanced evolutionary algorithms, the covariance matrix adaptation evolution strategy (CMA-ES) [26], to solve them. However, the responses of seismic structures are time-variant, so how to efficiently solve the interval bounds remains a challenge.

To efficiently conduct the interval uncertainty analysis for seismic structures, this paper first improves CMA-ES by low-discrepancy sequences (LDS). LDS is a sequence with excellent uniformity, which is widely applied in uncertainty quantification [27], experimental design [28], etc. There are various LDS, such as the Halton sequence [29], dynamic evolution sequence (DES) [30], and Sobol sequence [31]. Recently, Wu et al. [32] showed that integrating LDS into evolutionary algorithms improves offspring uniformity and solution space coverage, thereby enhancing efficiency. Among the tested LDS, DES exhibited the most significant performance gains. Therefore, this paper improves CMA-ES using DES and proposes DES-ES, whose efficiency is proven by benchmark functions. Meanwhile, inspired by the sequential simulation strategy, this paper further improves DES-ES by the response dependence and proposes a computational framework, DES-ES-SS. Numerical experiments demonstrate that DES-ES-SS enhances both efficiency and accuracy in interval uncertainty analysis for seismic structures under stationary seismic acceleration while improving computational efficiency by over an order of magnitude for non-stationary cases.

The rest of this paper is structured as follows. In section 2, the equation of motion of seismic structures and the typical PSD of seismic acceleration are introduced. MRIP for describing seismic acceleration is introduced in section 3. Section 4 presents the proposed framework, DES-ES-SS. Before concluding section 6, section 5 demonstrates the efficiency of DES-ES-SS by the numerical experiments.

## 2 Problem Overview

### 2.1 Equation of motion

The equation of motion of an *n*-degree-of-freedom seismic structure can be summarized as follows,

$$\boldsymbol{M}\ddot{\boldsymbol{X}}(t) + \boldsymbol{F}(\boldsymbol{X}, \dot{\boldsymbol{X}}, t) = -\boldsymbol{M}\boldsymbol{I}U(t), \tag{1}$$

where $\ddot{\boldsymbol{X}}(t)$, $\dot{\boldsymbol{X}}(t)$, and $\boldsymbol{X}(t)$ denote the acceleration, velocity, and displacement matrices; $\boldsymbol{M}$ denotes the mass matrix; $\boldsymbol{F}(\boldsymbol{X}, \dot{\boldsymbol{X}}, t)$ demotes the nonlinear internal-force vector of the structure; $\boldsymbol{I}$ represents the column vector with element 1; and $U(t)$ is the seismic acceleration.

In this paper, a 10-story frame structure is considered. The Bouc-Wen model [33] is adopted to describe the nonlinear behavior. Based on the rigid-floor assumption, the frame structure is simulated by a story-shear model, whose elastic modulus is $E = 2 \times 10^8$ MPa and the mass of each floor is $m = 250000$ kg. The structural damping adopts Rayleigh damping, and the first two damping ratios are taken as 0.05. Parameters of the Bouc-Wen model are valued by $A = 1$, $n = 1$, $\beta = 15 \text{ m}^{-1}$, $\gamma = 150 \text{ m}^{-1}$, $d_\nu = 0 \text{ m}^{-2}$, $d_\eta = 0 \text{ m}^{-2}$, $p = 0 \text{ m}^{-2}$, $q = 1$, $d_\psi = 1 \text{ m}^{-2}$, $\lambda = 1$, $\zeta_s = 0.95$, and $\psi = 1$ m.

### 2.2 Typical Power Spectrum Density Function of Seismic Acceleration

Seismic acceleration is often assumed to be a stationary or non-stationary stochastic process. For stationary seismic acceleration, its double-sided PSD is often assumed to be the Kanai-Tajimi spectrum [6, 7] or the Clough-Penzien spectrum [8], shown as follows,

$$\begin{cases} S_U^{\text{K-T}}(\omega) = \dfrac{\omega_g^4 + 4\zeta_g^2 \omega_g^2 \omega^2}{\left(\omega^2 - \omega_g^2\right)^2 + 4\zeta_g^2 \omega_g^2 \omega^2} S_0 \\ S_U^{\text{C-P}}(\omega) = \dfrac{\omega^4}{\left(\omega^2 - \omega_f^2\right)^2 + 4\zeta_f^2 \omega_f^2 \omega^2} S_{\ddot{X}}^{\text{K-T}}(\omega) \end{cases}, \qquad (2)$$

where $\omega_g$ and $\zeta_g$ denote the dominant frequency and damping ratio of site soil; $S_0$ denotes the spectral intensity of seismic acceleration, valued by $S_0 = 48.933 \text{ cm}^2/\text{s}^3$ according to Ref. [34]; $\omega_f$ and $\zeta_f$ denote the parameters of the filter hindering the low-frequency components of seismic acceleration, valued by $\omega_f = 0.1\omega_g$, $\zeta_f = \zeta_g$ according to Ref. [35].

For non-stationary seismic acceleration, its double-sided EPSD $S_{\tilde{U}}(\omega, t)$ is often defined as the product of PSD and a modulation function $A(\omega, t)$ [34], whose equation is as follows,

$$\begin{cases} S_{\tilde{U}}(\omega, t) = |A(\omega, t)|^2 S_U(\omega), \\ A(\omega, t) = \dfrac{e^{-at} - e^{-(c\omega+b)t}}{e^{-at^*} - e^{-(c\omega+b)t^*}}, \ t^* = \dfrac{\ln(c\omega+b) - \ln(a)}{c\omega + b - a}, \ \omega > 0, \ t > 0, \end{cases} \qquad (3)$$

## 3  Describe Seismic Acceleration with Incomplete Information

Section 2 briefly introduces the typical PSD and EPSD of seismic acceleration. However, since it is difficult to obtain samples of seismic acceleration, the uncertainty information is often incomplete. Therefore, it is appropriate to describe seismic acceleration by non-probabilistic methods [20]. Jiang et al. [21] used a convex model to describe time-variant uncertainty and proposed IP. On this basis, Wu et al. [23] proposed MRIP and its efficient construction method. This section will introduce how to use MRIP to describe seismic acceleration with incomplete information.

### 3.1 MRIP

For any instant $t_i$, $i = 1, 2, \cdots, N$, any possible values of $U(t)$ are enveloped in the interval $U^{\mathrm{I}}(t_i) = [U^{\mathrm{L}}(t_i), U^{\mathrm{U}}(t_i)]$, and $U^{\mathrm{I}}(t_i)$ can be called IP. The characteristic parameters of IP are the interval median $U^{\mathrm{m}}(t_i)$, the interval radius $U^{\mathrm{r}}(t_i)$, and the autocorrelation function $\rho(t_i, t_j)$, whose definitions are as follows,

$$U^{\mathrm{m}}(t_i) = \frac{U^{\mathrm{U}}(t_i) + U^{\mathrm{L}}(t_i)}{2}, \ U^{\mathrm{r}}(t_i) = \frac{U^{\mathrm{U}}(t_i) - U^{\mathrm{L}}(t_i)}{2}, \ \rho(t_i, t_j) = \frac{\mathrm{cov}(t_i, t_j)}{U^{\mathrm{r}}(t_i) U^{\mathrm{r}}(t_j)}, \quad (4)$$

where $\mathrm{cov}(t_i, t_j)$ is the covariance function, and the corresponding covariance matrix $\boldsymbol{C}$ is denoted by

$$\boldsymbol{C} = \begin{bmatrix} \mathrm{cov}(t_1, t_1) & \mathrm{cov}(t_1, t_2) & \cdots & \mathrm{cov}(t_1, t_N) \\ \mathrm{cov}(t_2, t_1) & \mathrm{cov}(t_2, t_2) & \cdots & \mathrm{cov}(t_2, t_N) \\ \vdots & \vdots & \ddots & \vdots \\ \mathrm{cov}(t_N, t_1) & \mathrm{cov}(t_N, t_2) & \cdots & \mathrm{cov}(t_N, t_N) \end{bmatrix}. \quad (5)$$

MRIP requires that the ellipsoid determined by $\boldsymbol{C}$ can enclose all the $N_{\mathrm{s}}$ measured samples $\boldsymbol{U}_k = (U_k(t_1), U_k(t_2), \cdots, U_k(t_N))^{\mathrm{T}}$, $k = 1, 2, \cdots, N_{\mathrm{s}}$, where $U_k(t_i)$ is the $k$-th measured seismic acceleration at instant $t_i$. So, it satisfies,

$$(\boldsymbol{U}_k - \boldsymbol{o})^{\mathrm{T}} \boldsymbol{C}^{-1} (\boldsymbol{U}_k - \boldsymbol{o}) \leq 1, \ k = 1, 2, \cdots, N_{\mathrm{s}}, \quad (6)$$

where $\boldsymbol{o} = (U^{\mathrm{m}}(t_1), U^{\mathrm{m}}(t_2), \cdots, U^{\mathrm{m}}(t_N))^{\mathrm{T}}$ is the center of the ellipsoid. Meanwhile, MRIP requires minimizing the trace of $\boldsymbol{C}$ [23]. Therefore, constructing MRIP can be summarized as follows,

$$\begin{cases} \min \ \mathrm{trace}(\boldsymbol{C}) \\ \mathrm{s.t.} \ (\boldsymbol{U}_k - \boldsymbol{o})^{\mathrm{T}} \boldsymbol{C}^{-1} (\boldsymbol{U}_k - \boldsymbol{o}) \leq 1, \ k = 1, 2, \cdots, M, \\ \boldsymbol{C} \text{ is a symmetric positive definite matrix} \end{cases} \quad (7)$$

For the stationary IP (SIP), it is required [21],

$$U^{\mathrm{m}}(t_i) = U^{\mathrm{m}}, \ U^{\mathrm{r}}(t_i) = U^{\mathrm{r}}, \ \rho(t_i, t_j) = \rho(\tau), \ \rho(t_i, t_j) = \rho(\tau = |t_i - t_j|). \quad (8)$$

Therefore, constructing the minimum interval radius-based SIP (MRSIP) can be summarized as follows [23],

$$\begin{cases} \min \ U^{\mathrm{r}} \\ \mathrm{s.t.} \ (\boldsymbol{U}_k - \boldsymbol{o})^{\mathrm{T}} \boldsymbol{C}^{-1} (\boldsymbol{U}_k - \boldsymbol{o}) \leq 1, \ k = 1, 2, \cdots, M \\ \boldsymbol{C} \text{ is a symmetric positive definite Toeplitz matrix} \end{cases}. \quad (9)$$

### 3.2 Interval K-L expansion

To perform fast sampling for IP, Ni and Jiang [22] proposed the interval K-L expansion,

$$U^s(t) = U^m(t_i) + \sum_{j=1}^{\infty} U^r(t_i)\sqrt{\lambda_j}\varphi_j(t)\theta_j, \tag{10}$$

where $\sum_{j=1}^{\infty}\theta_j^2 \leq 1$, $j = 1, 2, \cdots$; $\lambda_j$ and $\varphi_j(t)$ are the eigenvalue and eigenfunction of $\rho(t_i, t_j)$ [22]. Moreover, the series terms can be sorted in descending order according to the magnitude of $\lambda_j$ and then need to be truncated,

$$\tilde{U}^s(t) = U^m(t_i) + \sum_{j=1}^{M} U^r(t_i)\sqrt{\lambda_j}\varphi_j(t)\theta_j, \tag{11}$$

where $\sum_{j=1}^{M}\theta_j^2 \leq 1$ represents points within an *M*-dimensional hypersphere. Furthermore, the steps for using MRIP to describe seismic acceleration are as follows:

A. Obtain the sample points $\boldsymbol{U}_k = (U_k(t_1), U_k(t_2), \cdots, U_k(t_N))^T$ from the measured seismic acceleration samples, and construct MRIP by solving the optimization equation (7) or (9).

B. Randomly draw samples $\tilde{\boldsymbol{\theta}}$, and obtain $\boldsymbol{\theta}$ within the *M*-dimensional hypersphere through the Nataf transformation.

C. Map $\boldsymbol{\theta}$ to the sample function $\tilde{U}^s(t)$ of the seismic acceleration through equation (11).

It is worth noting that MRIP does not involve probability information. Therefore, when using MRIP to describe seismic acceleration, only the envelope interval of the response needs to be solved.

## 4 The Proposed Method

In section 3, we introduced how to use MRIP to describe seismic acceleration, and then conduct interval uncertainty analysis on the seismic structure. That is to solve the extreme value of the response at each instant as shown in the following optimization equation,

$$\begin{cases} \min\ y(t_i) \\ \max\ y(t_i) \end{cases}, i = 1, 2, \cdots, N, \tag{12}$$

where $y(t_i)$ represents the value of the response at instant $t_i$. Wu et al. [25] utilized CMA-ES to conduct the interval uncertainty analysis. However, the response they considered is time-invariant, resulting in fewer optimization problems. But for the seismic structure, the response is often time-variant, resulting in a huge amount of computation. On these bases, this paper improves CMA-ES using LDS and further optimizes the solution strategy to propose a calculation framework, DES-ES-SS.

### 4.1 CMA-ES

CMA-ES [26] is based on a multivariate normal distribution sequence to obtain the individuals. For CMA-ES, whose number of offspring is $\lambda$, the $i$-th individual at the $g$-th generation is generated as the following equation,

$$x_{g,i} = m_{g-1} + \sigma_{g-1} B_{g-1} \left( d_{g-1} \circ \Phi^{-1}\left(\varepsilon_{g,i}\right)\right), \quad i = 1, 2, \ldots, \lambda, \tag{13}$$

where $m_{g-1}$ is the $D \times 1$ mean value of the search distribution; $\sigma_{g-1}$ is the step-size; $C_{g-1}$ is the $D \times D$ matrix; $B_{g-1}$ is a $D \times D$ matrix formed by the eigenvectors of $C_{g-1}$; $d_{g-1}$ is a $D \times 1$ column vector created by the eigenvalues of $C_{g-1}$; $\Phi^{-1}(\cdot)$ is the inverse function of the probability distribution function of standard normal distribution; $\varepsilon_{g,i}$ is a $D \times 1$ random sequence within the interval $[0,1]$. See Ref. [26] for the detailed calculation steps.

### 4.2 Improvement of CMA-ES

Although CMA-ES has advantages such as strong search capabilities, the uniformity of its offspring is poor, and it cannot cover the solution space well. Wu et al. [32] demonstrated that for evolutionary algorithms, replacing uniformly distributed random variables with LDS can make the offspring more uniform and better cover the solution space, thus improving the solution efficiency. Meanwhile, DES [30] has the most obvious effect on improving efficiency. Therefore, this paper improves CMA-ES by DES and proposes an efficient evolutionary algorithm, DES-ES.

#### 4.2.1 Dynamic evolution sequence

The static solutions of multi-body systems often exhibit good uniformity. Inspired by this, Wu et al. [30] proposed DES. Assume that all samples are within the $D$-dimensional unit cube, and each sample is regarded as a star with a mass of $m$. The coordinate of the $i$-th star is $x_i = (x_{i1}, \cdots, x_{iD})$, and there are interaction forces between points. The Lagrangian equations of the stars are,

$$\begin{cases} S = \int_0^t L \mathrm{d}\tau \\ L = \dfrac{1}{2} m \sum_{i=1}^{N} \sum_{k=1}^{D} \dot{x}_{ik}^2 - G \left( \sum_{1 \le i < j \le N} \dfrac{1}{d_{q,ij}^p} \right)^{\frac{1}{p}} \end{cases}, \tag{14}$$

where $G = 1$ is the generalized gravitational constant; $q$ and $p$ are control parameters; and $d_{q,ij}$ is the generalized distance. The values can be found in Ref. [30]. According to Hamilton's variational principle [36-39], it can be obtained that,

$$m\ddot{x}_{ik} + c\dot{x}_{ik} + f_{ik} = 0,\ 1 \le i \le N,\ 1 \le k \le D, \tag{15}$$

where $c$ is an artificial damping coefficient. When the system is static, the sequence $X_{D,N} = \{x_1, x_2, \cdots, x_N\}$ composed of the obtained galaxy coordinates $x_i$ can be regarded as LDS.

### 4.2.2 DES-ES

Assumed a DES of $D \times \lambda$ size, called $P$, is obtained by equations (14)-(15),

$$P = \begin{bmatrix} p_{11} & p_{12} & \cdots & p_{1\lambda} \\ p_{21} & p_{22} & \cdots & p_{2\lambda} \\ \vdots & \vdots & \ddots & \vdots \\ p_{D1} & p_{D2} & \cdots & p_{D\lambda} \end{bmatrix} = \begin{bmatrix} p_{1,:} \\ p_{2,:} \\ \vdots \\ p_{D,:} \end{bmatrix}. \tag{16}$$

Then, each column vector of $P$ can be used to replace the random vector $\varepsilon_{g,i}$ in equation (13), making the offspring cover the solution space more uniformly. However, it causes CMA-ES to lose the randomness [40]. To enable CMA-ES to increase uniformity with randomness, a method called dimensional random rearranging (DRR) is proposed. The specific steps of DRR are as follows:

A. Let $I_0 := \{1, 2, ..., D\}$ and randomly select a number $\Phi_{1,1}$ from $I_0$;

B. Let $j = 1, 2, ..., D$ and do the following loop,

    a) Put the $\Phi_{1,j}$-th row elements in $P$ into the $j$-th row elements of $P_\Phi$;

    b) $I_j = I_{j-1} - \Phi_{1,j}$. This operation denotes, removing the selected number $\Phi_{1,j}$ from the set $I_{j-1}$ to generate the new set $I_j$;

    c) Randomly select a number $\Phi_{1,j}$ from the set $I_j$.

Finally, a new sequence $P_\Phi$ with randomness is generated by the above operations,

$$P_{\Phi_1} = \begin{bmatrix} p_{\phi_{1,1},1} & p_{\phi_{1,1},2} & \cdots & p_{\phi_{1,1},\lambda} \\ p_{\phi_{1,2},1} & p_{\phi_{1,2},2} & \cdots & p_{\phi_{1,2},\lambda} \\ \vdots & \vdots & \ddots & \vdots \\ p_{\phi_{1,D},1} & p_{\phi_{1,D},2} & \cdots & p_{\phi_{1,D},\lambda} \end{bmatrix} = \begin{bmatrix} p_{\phi_{1,1},:} \\ p_{\phi_{1,2},:} \\ \vdots \\ p_{\phi_{1,D},:} \end{bmatrix}. \tag{17}$$

It can be proved that the uniformity of $P_\Phi$ is the same as that of $P$. There are many commonly used discrepancies to measure uniformity, such as $L_2$ discrepancy, MD discrepancy, and WD discrepancy [41]. For example, the $L_2$ discrepancy of $P$ and $P_\Phi$ are expressed by,

$$D_{L_2}(P)^2 = \left(\frac{1}{3}\right)^D - \frac{2}{D}\sum_{j=1}^{\lambda}\prod_{i=1}^{D}\left(\frac{1-p_{ij}^2}{2}\right) + \frac{1}{D^2}\sum_{j,l=1}^{\lambda}\prod_{i=1}^{D}\left[1-\max\left(p_{ij}, p_{il}\right)\right], \tag{18}$$

and,

$$D_{L_2}(P_{\Phi_g})^2 = \left(\frac{1}{3}\right)^D - \frac{2}{D}\sum_{j=1}^{\lambda}\prod_{i=1}^{D}\left(\frac{1-p_{\phi_{i,g},j}^2}{2}\right) + \frac{1}{D^2}\sum_{j,l=1}^{\lambda}\prod_{i=1}^{D}\left[1-\max\left(p_{\phi_{i,g},j}, p_{\phi_{i,g},l}\right)\right]. \tag{19}$$

respectively. It can be derived that,

$$D_{L_2}(P)^2 = D_{L_2}(P_{\Phi_g})^2. \tag{20}$$

Similarly, it can be proved that DRR does not change the other discrepancies. Therefore, DRR does not change the uniformity of the original DES. So, DRR the DES $P$ $G-1$ times to generate $G-1$ different LDSs of $D \times \lambda$ size, named $P_{\Phi_1}, P_{\Phi_2}, ..., P_{\Phi_{G-1}}$. Each column vector of $P$ and $P_{\Phi_g}$ can be used to replace the random vector $\varepsilon_{g,i}$ in equation (13), making the offspring cover the solution space more uniformly to enhance the efficiency. The more efficient CMA-ES based on DES is called DES-ES.

### 4.3 Interval Uncertainty Analysis Framework of Seismic Structures Based on DES-ES

For evolutionary algorithms, the closer the first-generation offspring $x_{1,i}$ is to the optimal solution, the faster the algorithm can find the optimal solution. Therefore, for DES-ES, the closer the initially given mean value $m_0$ is to the optimal solution, the higher its solution efficiency. Ni et al. [24] proposed the sequential simulation strategy, which improves the efficiency of interval uncertainty analysis by taking advantage of the dependency of the responses. Inspired by the sequential simulation strategy [24], this section will further enhance efficiency.

For seismic structures, the responses should be continuous. Thus, it can be assumed that if a randomly sampled $\tilde{\theta}$ generates a seismic acceleration through the interval K-L expansion such that the considered response reaches the extreme value at the instant $t_i$, then this seismic acceleration should cause the considered response at the instant $t_{i+1}$ to take a value near the extreme value. Therefore, it can be considered that the sample $\tilde{\theta}^{\max t_i}$ that makes the considered response reach the extreme value at instant $t_i$ is in the vicinity of the sample $\tilde{\theta}^{\max t_{i+1}}$ that makes the considered response reach the extreme value at instant $t_{i+1}$. In summary, the optimal solution $x^{gbest\ t_i}$ obtained by DES-ES for solving min $y(t_i)$ can be used as the initial mean value $m_0$ for DES-ES to solve min $y(t_{i+1})$, which makes the first-generation offspring $x_{1,i}$ closer to the optimal solution, thereby improving the efficiency of interval uncertainty analysis for seismic structures. Where, $y(t_i)$ represents the value of the considered response at the instant $t_{i+1}$. The interval uncertainty analysis framework for seismic structures, which is conducted according to the aforementioned method, is called DES-ES-SS. Taking the solution of the lower bound of the response envelope interval as an example, the specific steps of DES-ES-SS are shown in Fig. 1.

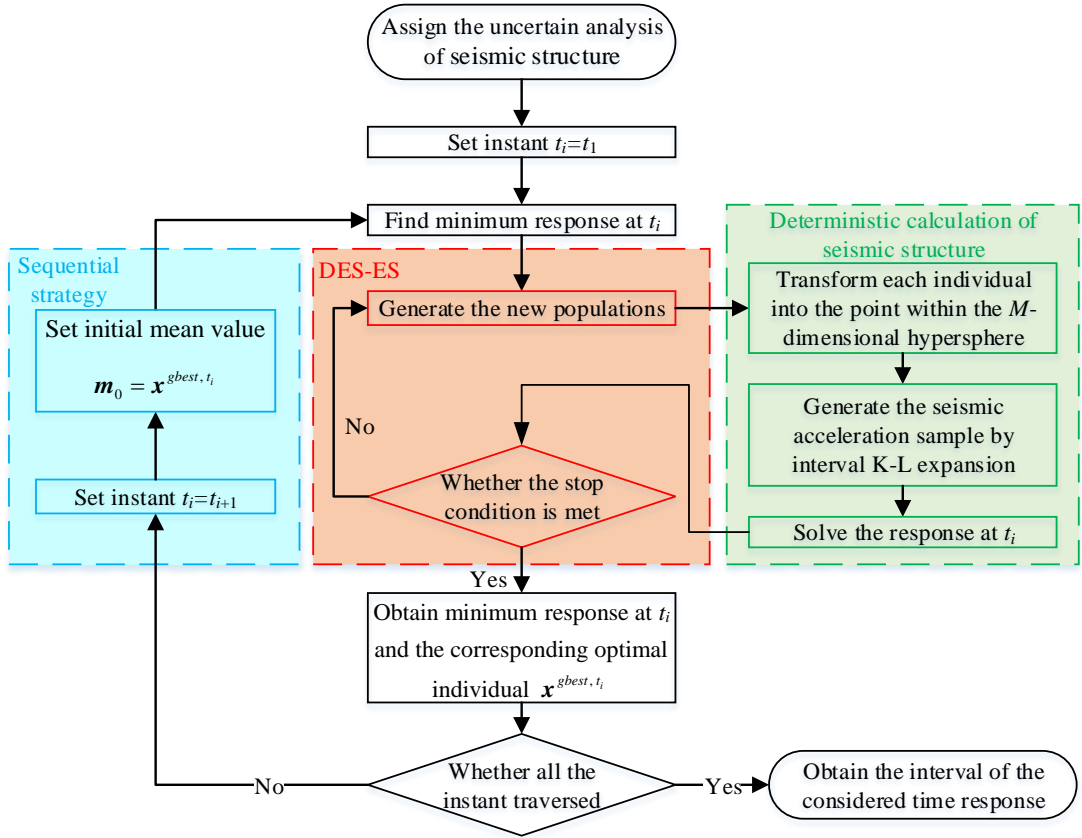

**Fig. 1 Solution flow chart of DES-ES-SS.**

## 5 Numerical Experiments

### 5.1 Verification of efficiency of DES-ES

First, ten benchmark functions [42] are selected to test the performance of DES-ES. When relative error $e$ is less than error tolerance $\varepsilon_{tol}$, the algorithm is considered to converge. Due to the accuracy of the algorithm being affected by dimension, for the 10-dimensional problems the $\varepsilon_{tol} = 5\%$ and $1\%$, and for the 30-dimensional and 50-dimensional problems the $\varepsilon_{tol} = 20\%$ and $10\%$. Both CMA-ES and DES-ES are utilized to solve this set of functions respectively. The convergent number of iterations regarding the residual $\varepsilon_{tol}$ is shown in Table 1.

As shown in Table 1, for the benchmark functions, the convergence speed of DES-ES is almost always faster than that of CMA-ES. Therefore, it can be considered that DES-ES can significantly improve efficiency.

Table 1 The number of iterations required for convergence.

| Function name | 10-dimension | | | | 30-dimension | | | | 50-dimension | | | |
|---|---|---|---|---|---|---|---|---|---|---|---|---|
| | $\varepsilon_{tol}$=5% | | $\varepsilon_{tol}$=1% | | $\varepsilon_{tol}$=20% | | $\varepsilon_{tol}$=10% | | $\varepsilon_{tol}$=20% | | $\varepsilon_{tol}$=10% | |
| | CMA-ES | DES-ES | CMA-ES | DES-ES | CMA-ES | DES-ES | CMA-ES | DES-ES | CMA-ES | DES-ES | CMA-ES | DES-ES |
| Shifted and rotated Bent Cigar function | 100 | 84 | 105 | 89 | 305 | 192 | 310 | 194 | 459 | 297 | 464 | 300 |
| Shifted and rotated Zakharov function | 45 | 37 | 51 | 41 | 201 | 155 | 207 | 160 | 459 | 363 | 466 | 372 |
| Shifted and rotated Rosenbrock's function | 21 | 19 | 79 | 70 | 53 | 35 | 66 | 43 | 86 | 54 | 452 | 371 |
| Shifted and rotated Rastrigin's function | 58 | 38 | 91 | 50 | 146 | 59 | 180 | 66 | 220 | 73 | 248 | 79 |
| Shifted and rotated Expaned Schaffer function | 12 | 11 | 27 | 23 | 3 | 2 | 22 | 15 | 13 | 8 | 30 | 18 |
| Shifted and rotated Lunacek Bi-Ratrigin's function | 75 | 46 | - | - | 198 | 67 | 265 | 78 | 343 | 85 | 516 | 99 |
| Shifted and rotated Non-continuous Rastrigin's function | 24 | 19 | 80 | 45 | 109 | 46 | 156 | 59 | 204 | 67 | 245 | 75 |
| Shifted and rotated Lvey function | 21 | 17 | 28 | 23 | 54 | 33 | 61 | 37 | 85 | 48 | 93 | 52 |
| Zakharov, Rosenbrock, Rastrigin hybrid function | 30 | 22 | 50 | 40 | 123 | 68 | 155 | 132 | 357 | 182 | - | 775 |
| Katsuura; Ackley; Rastrigin; Schaffer; Modified Schwefel hybrid function | 39 | 27 | 115 | - | 150 | 46 | 220 | 72 | 378 | 83 | 464 | 116 |

### 5.2 Uncertainty Analysis of Frame Structure under Stationary Seismic Acceleration

For the frame structure introduced in section 2.1, assume it is subjected to stationary seismic acceleration. A total of 20 artificial seismic accelerations, whose PSD is the Clough-Penzien spectrum, where $\omega_g = 5\pi$ rad/s and $\zeta_g = 0.6$, are generated as the original samples, and the stationary MRIP is constructed by solving the equation (9) to describe the seismic acceleration.

The interval upper and lower bounds of the displacement, velocity, and acceleration of the top floor within 10 s with a step size of 0.05 s are solved by MCS (with $10^6$ samples), CMA-ES, and DES-ES-SS. The time-history response envelope curves obtained by each method are shown in Fig. 2. Moreover, MCS takes 46941.8678 s, CMA-ES takes 9210.9121 s, and DES-ES-SS takes 7871.9010 s.

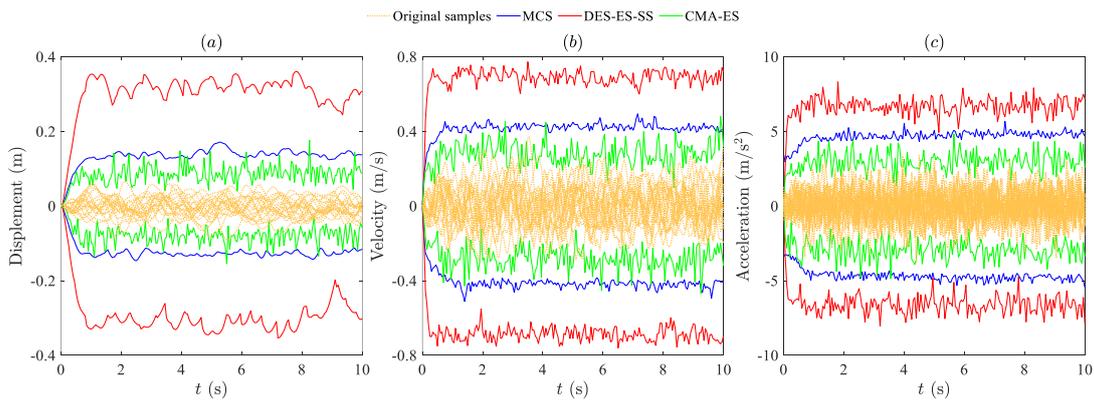

**Fig. 2 Time-history response envelope curves of the frame structure under seismic acceleration described by MRSIP: (a) Displacement; (b) Velocity; (c) Acceleration.**

As can be seen from Fig. 2, the time-history response envelope curves obtained by the three methods can envelope the responses generated under the original samples. Meanwhile, DES-ES-SS can identify the most extreme cases. Therefore, the solution obtained by DES-ES-SS is closest to the actual solution That is, DES-ES-SS has the highest accuracy, followed by MCS, and CMA-ES has the lowest accuracy. Furthermore, by comparing the computation times of different methods, it can be found that the computation time of DES-ES-SS is only 16.77% of that of MCS. Therefore, it can be considered that, compared with MCS, DES-ES-SS can significantly improve the accuracy while increasing the efficiency by more than 80%.

**5.3 Uncertainty Analysis of Frame Structure under Non-Stationary Seismic Acceleration**

Next, it is amused the frame structure is subjected to non-stationary seismic acceleration. A total of 20 artificial seismic accelerations, whose EPSD is the product of the Clough-Penzien spectrum and the modulation function defined in Equation (3), where $a = 0.05 \text{ s}^{-1}$, $b = a + 0.02$, and $c = 0.01$, are generated as the original samples. The MRIP is constructed by solving Equation (7) to describe the seismic acceleration.

The interval upper and lower bounds of the displacement, velocity, and acceleration of the top floor within 10 s with a step size of 0.05 s are solved by MCS (with $10^6$ samples), CMA-ES, and DES-ES-SS. The time-history response envelope curves obtained by each method are shown in Fig. 3. Moreover, MCS takes 187818.4268 s, CMA-ES takes 7233.2513 s, and DES-ES-SS takes 7231.6408 s.

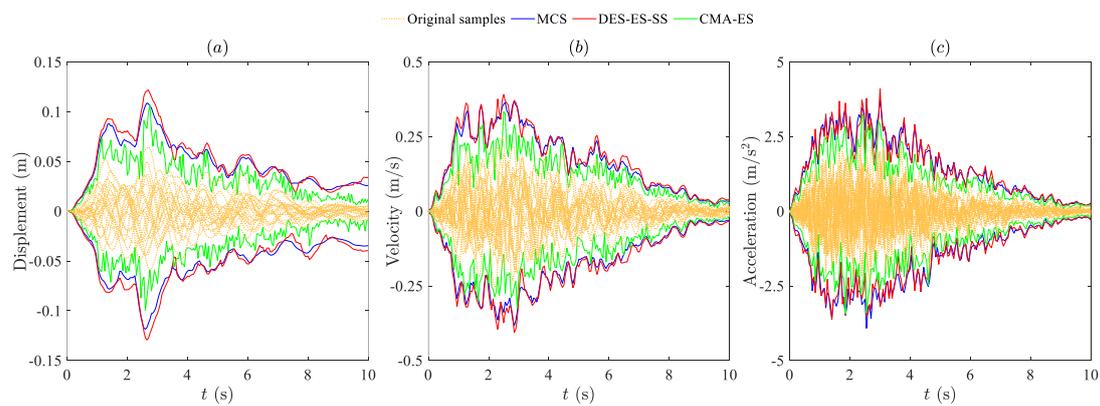

**Fig. 3 Time-history response envelope curves of the frame structure under non-stationary seismic acceleration described by MRIP: (a) Displacement; (b) Velocity; (c) Acceleration.**

According to Fig. 3, the time-history response envelope curves obtained by MCS and DES-ES-SS can both envelope the responses generated under the action of the original seismic accelerations. The time-history response envelope curves obtained by CMA-ES can envelop the responses under the original samples at most instant. Meanwhile, The time-history response envelope curves obtained by DES-ES-SS and MCS are similar. So, the accuracy of the DES-ES-SS and MCS is similar, while that of the CMA-ES is the worst. Furthermore, by comparing the computational times of different methods, it can be found that the computational time of DES-ES-SS is only 3.85% of that of MCS. Therefore, for the uncertainty analysis of frame structures under non-stationary seismic acceleration described by MRIP, it can be considered that DES-ES-SS can improve computational efficiency by more than one order of magnitude while maintaining accuracy.

## 6 Conclusion

Non-probabilistic methods are appropriate to describe the seismic acceleration with incomplete uncertainty information. So, this paper uses MRIP, which is based on the non-probabilistic convex model, to describe seismic acceleration and conducts interval uncertainty analysis. However, traditional MCS requires extensive sampling for accuracy, leading to low efficiency.

To address this issue, this paper enhances CMA-ES with DES, proposing DES-ES for higher efficiency, and leverages the dependence of the response to develop a calculational framework, DES-ES-SS. Then, tests on ten benchmark functions confirm the efficiency of DES-ES. Subsequently, interval uncertainty analyses are conducted for seismic structures under stationary and non-stationary seismic acceleration. Results show that for stationary seismic acceleration, DES-ES-SS outperforms MCS in both efficiency and accuracy, while for non-stationary cases, DES-ES-SS achieves over tenfold efficiency gains compared to MCS. Therefore, it can be considered that DES-ES-SS can significantly improve the efficiency of interval uncertainty analysis for seismic structures under seismic acceleration described by MRIP.

In the future, we will extend DES-ES-SS to the uncertainty analysis of more time-variant uncertainty systems with incomplete uncertainty information.